\documentclass[conference]{IEEEtran} 

\usepackage{amssymb}

\usepackage{lineno}
\usepackage{epsfig}
\usepackage[utf8]{inputenc}
\usepackage{times}
\usepackage{graphicx}
\usepackage[figuresright]{rotating}
\usepackage{float}
\usepackage{amssymb}
\usepackage{amsmath}
\usepackage{url}
\usepackage{hyperref}
\usepackage{soul}
\usepackage{wrapfig}
\floatname{algorithm}{Algorithm}
\usepackage[usenames]{color}
\usepackage{flushend}
\usepackage[nolist]{acronym}
\usepackage{lmodern,textcomp}
\usepackage{multirow}
\usepackage{colortbl}
\makeatletter
\newcommand{\thickhline}{%
    \noalign {\ifnum 0=`}\fi \hrule height 1.5pt
    \futurelet \reserved@a \@xhline
}
\newcolumntype{"}{@{\hskip\tabcolsep\vrule width 1pt\hskip\tabcolsep}}
\makeatother
\newcolumntype{?}{!{\vrule width 1.5pt}}
\usepackage[table]{xcolor} 
\usepackage{etoolbox}

\usepackage{booktabs}

\begin{document}
\newcommand{\squeezeupppp}{\vspace{-8 mm}}
\newcommand{\squeezeuppp}{\vspace{-6 mm}}
\newcommand{\squeezeupp}{\vspace{-5 mm}}
\newcommand{\squeezeup}{\vspace{-3 mm}}
\newcommand{\squeezeu}{\vspace{-2 mm}}
\newcommand{\squeeze}{\vspace{-1 mm}}
\newcommand{\squeez}{\vspace{-.5 mm}}

\begin{acronym}
\acro{ACK}{Acknowledgment}
\acro{ACN}{Availability Confirmation}
\acro{AH}{Aerial HELPER}
\acro{AODV}{Ad hoc On-Demand Distance Vector}
\acro{AP}{Access Point}
\acro{API}{Application Programming Interface}
\acro{BE}{Backoff Exponent}
\acro{BER}{Bit Error Rate}
\acro{BI}{Beacon Interval}
\acro{CAP}{Contention Access Period}
\acro{CAD}{Channel Activity Detection}
\acro{CC}{Convolutional Coding}
\acro{CCA}{Clear Channel Assessment}
\acro{CDMA}{Code Division Multiple Access}
\acro{CFP}{Contention Free Period}
\acro{CSMA/CA} {Carrier Sense Multiple Access/Collision Avoidance}
\acro{CSMA/CD} {Carrier Sense Multiple Access/Collision Detection}
\acro{CTS}{Clear-to-send}
\acro{DMT}{Discrete Multi-Tones}
\acro{D2D}{Device-to-Device}
\acro{DoA} {Direction of Arrival}
\acro{DoD}{Department Of Defense}
\acro{ECC}{Error-Correction Code}
\acro{EMI}{Electromagnetic Interference}
\acro{ER}{Emergency Responder}
\acro{ERC}{Emergency Response Center}
\acro{EU}{End User}
\acro{FCC}{Federal Communications Commission}
\acro{FEC}{Forward Error Correction}
\acro{FEMA}{Federal Emergency Management Agency}
\acro{FSM}{Finite State Machine}
\acro{FSO}{Free Space Optics}
\acro{FOV}{Field Of View}
\acro{GPS}{Global Positioning System}
\acro{GTS}{Guaranteed Time Slots}
\acro{HTL}{Hop-To-Live}
\acro{IM/DD}{Intensity-Modulation Direct-Detection}
\acro{IoT}{Internet of Things}
\acro{ISM}{Industrial, Scientific and Medical}
\acro{ISR}{ Intelligence, Surveillance, and Reconnaissance}
\acro{LOS}{Line of Sight}
\acro{LPI/LPD}{Lower Probability of Intercept/Lower Probability of Detection}
\acro{LTE}{Long-Term Evolution}
\acro{MAC}{Medium Access Control}
\acro{MANET}{Mobile Ad Hoc Network}
\acro{MH}{Mobile HELPER}
\acro{MUI}{Multi-User Interference}
\acro{NAV}{Network Allocation Vector}
\acro{NB}{Number of Backoffs}
\acro{ND}{Network Discovery}
\acro{MC-CDMA}{ Multi-carrier \ac{CDMA}}
\acro{NLOS}{non-Line Of Sight}
\acro{NRL}{Naval Research Labs}
\acro{OAI}{Optimization Assisting Information}
\acro{OCDMA}{Optical Code-Division Multiple Access}
\acro{OFDM}{Orthogonal Frequency Division Multiplexing}
\acro{OFDMA}{Orthogonal Frequency Division Multiple Access}
\acro{OLSR}{Optimized Link State Routing}
\acro{PHR}{PHY Header}
\acro{PHY}{Physical} 
\acro{QoS}{Quality of Service}
\acro{RA}{Random Access}
\acro{RES}{Reserve Sectors}
\acro{RF}{Radio Frequency}
\acro{RPI}{Raspberry Pi}
\acro{RSSI}{Received Signal Strength Indication}
\acro{SH}{Static HELPER}
\acro{SNR}{Signal-to-Noise Ratio}
\acro{SWaP}{(Size, Weight, and Power)}
\acro{TDD}{Time Division Duplex}
\acro{TDMA}{Time Division Multiple Access}
\acro{TIA}{Telecommunication Industry Association}
\acro{USRP}{Universal Software Radio Peripheral}
\acro{UVC}{Ultraviolet Communication}
\acro{VHF}{Very High Frequency}
\acro{VLC}{Visible Light Communication}
\acro{V2I} {Vehicle to Infrastructure}
\acro{V2V} {Vehicle to Vehicle}
\acro{Web App}{Website Application}
\acro{WEA}{Wireless Emergency Alerts}
\acro{WiFi}{Wireless Fidelity}
\acro{WiMAX}{Worldwide Interoperability for Microwave Access}
\acro{WSN}{Wireless Sensor Network}
\end{acronym}



\title{Jam-Guard: Low-Cost, Hand-held Device for First Responders to Detect and Localize Jammers}





\author{\IEEEauthorblockN{Anu Jagannath}

\IEEEauthorblockA{Marconi-Rosenblatt AI/ML Innovation Laboratory, \\
ANDRO Computational Solutions, LLC, \\
Rome, NY, USA, 13440\\
E-mail: ajagannath@androcs.com 
}
\and
\IEEEauthorblockN{Jithin Jagannath}
\IEEEauthorblockA{Marconi-Rosenblatt AI/ML Innovation Laboratory, \\
ANDRO Computational Solutions, LLC, \\
Rome, NY, USA, 13440\\
E-mail: jjagannath@androcs.com}

}
\maketitle

\begin{abstract}
Intentional and unintentional interferences collectively referred to as Radio Frequency Interference (RFI) result in severe security threat to the public safety, first responder emergency rescue and military missions. Such RFI if not detected and localized can disrupt the wireless communication which forms the backbone of first responder and military operations. The prime objective of this work was to design and prototype a RFI detection and localization device, Jam-Guard that significantly outperforms traditional approaches in real-life deployment yet be computationally feasible to be developed as a low SWaC (size, weight, and cost) device. The proposed device employs a unique combination of robust parallel detection algorithms based on Kurtosis and FRactional Fourier Transform (FRFT) with Golden Section Search algorithm to rapidly detect RFI that affects critical communication signals. The localization scheme is designed to leverage the FRFT output from the detection phase to ease the computational load. We give a detailed account of the device prototyping and evaluation. To demonstrate the efficacy of the proposed detection approach as opposed to conventional energy detection (employed in several commercial interference detectors), we compare the two schemes on test and target platforms. The proposed scheme depicted significant improvement ($\mathbf{\sim 40\;dB}$) in detection probability both in preliminary implementation and target prototype experiments.
\end{abstract}


\section{Introduction}\label{sec:intro}
The recent surge of cheap portable jammers and their ease of availability to commoners from illicit online stores poses a severe threat to national security. Most of these jammers block the Global Positioning System (GPS), Cellular, LoJack, WiFi, public safety and law enforcement communication bands (disrupting 9-1-1 calls) and so on.  We will collectively refer to such intentional interferences and unintentional emissions from faulty electronics as radio frequency interference (RFI). Military as well as civilians rely heavily on GPS and cellular bands for position, navigation and timing (PNT) data and wireless communication. Military's significant reliance on GPS for PNT information also makes it a plausible target. For example, during tactical missions, the Air Force Space command might position the GPS satellites over the target area for precise strikes. Similar missions would require precise PNT information and undeterred communication to prevent any civilian casualties. During such events, detecting, localizing and mitigating any such jammers in the vicinity is key to the mission's success and to prevent casualties. 

There have been several such reports of illegal jammer use interfering with airport ground-based augmentation system (GBAS), maritime navigation, LoJack car security systems, cell phones etc. One such incident is the interference caused to the GBAS at the Newark Liberty International Airport in 2012 by the GPS jammer used by a New Jersey truck driver \cite{newark}. The truck driver used the jammer for personal privacy to prevent his employer from tracking the company-operated truck's whereabouts using the GPS tracking system installed on the truck. Another similar incident took place in France in 2017, after a 50-yr old man left an activated GPS jammer in his car and forgot to switch it off prior to boarding a flight from the Nantes Airport leading to severe communication disruption causing delayed flights \cite{nantes}. 
Several such jammer related incidents are reported worldwide and are a significant risk to public safety and a potential tool for terrorist activities. 
First responders require seamless communication to effectively carry out rescue missions and to communicate back to the base to apprise them of the current situation and dispatch officers to the site. Any communication disruption amidst such rescue missions can be fatal to both the victims and the first responders. Accordingly, the capability to identify the cause of communication disruption and localizing the source of RFI will be a significant step towards jammer mitigation. Therefore, in this work, we will design and prototype a low cost, portable device capable of detecting and localizing RFI in an efficient manner. The portability of the device is considered as a prime factor in our design constraint to allow first responders to carry it around with ease and install on the patrol vehicles. The low cost is another notable characteristic to allow rapid and affordable customer acquisition. This work significantly expands upon our previous work \cite{idad} which included the initial prototyping stages and preliminary results. 


\section{Related Works}\label{sec:related}
Signal detection and characterization is a well researched topic \cite{fftfreq, fft, sonarFRFT,KurtFRFT,cyclosigdet, Jagannath19MLBook} through theoretical approaches, simulations and few experimental analysis \cite{Jagannath_FACT, exp1,sentinel, Jagannath_ICC, Jagannath17CCWC}.  A frame-based Fast Fourier Transform (FFT) approach is proposed in \cite{fftfreq} to detect chirp signal and to estimate its chirp rate. This approach estimates the peak frequency (frequency bin with the maximum magnitude) across consecutive FFT frames and uses the peak frequency difference to estimate chirp rate and as the indicator to chirp signal presence. The algorithm is implemented and verified by simulations and by measured data.  An overlapped FFT based energy detection technique is mentioned in \cite{fft}. The authors propose window functions that reduces the correlation between adjacent FFT frames and numerically validate that the window function obtained by upsampling Hamming window improves the detection performance. All of these works were evaluated by simulations.  A framework for automatic modulation classification (FACT) is proposed and validated in \cite{Jagannath_FACT} which performs a conventional energy detection followed by  modulation classification. 
The effectiveness of FACT was validated on a five USRP testbed. 
Since their focus was to propose a framework for signal detection and classification, the authors only considered narrowband signals and use a traditional FFT based energy detector for signal detection.  \cite{exp1} propose a FFT and kurtosis based approach to detect signal presence whereby the kurtosis of FFT of the signal forms the test statistic. Authors implemented the proposed approach on software defined radio (SDR) using LabView. Furthermore, most commercial RFI detectors like CTL3520, CTL3510 predominantly depend on conventional energy detector to identify the presence of RFI. CTL3520 and CTL3510 are the commercial outcomes of the UK based SENTINEL project \cite{sentinel}. The SENTINEL system adopts a signal-to-noise-ratio (SNR) and FFT-based algorithm to perform RFI detection. The main shortcoming of FFT-based approaches discussed above is its poor performance at low SNR. In these cases, other characteristics of the signal has to be extracted to ensure high probability of detection. 

Accordingly, the authors propose a joint signal signal detection and classification approach in \cite{cyclosigdet}. The authors adopt a cyclostationary approach to determine the spectral coherence density of the signals to detect followed by a Hidden Markov Model to classify them. Such cyclostationary technique can be very effective at lower SNRs but incurs high computational complexity and requires larger samples to accurately detect and classify. These computational and processing constraints necessitate the need for faster expensive processors and would be harder to extend to low cost, portable platforms. Unlike the cyclostationary approaches, we focus on FRactional Fourier Transform (FRFT) which is implemented to be computationally efficient while providing accurate detection with  fewer samples as will be explained in the upcoming sections. 

 In \cite{sonarFRFT}, an FRFT based approach to detect sonar echoes of a desired chirp signal of interest from cluttered background is introduced. The FRFT at the matched order of the desired chirp would appear as a tone. Everywhere except the tone is notched out to suppress the noise from the received signal samples. The authors verified the algorithmic correctness of proposed approach by simulations. Similarly, the authors propose to use FRFT in conjunction with kurtosis to detect chirp signals in \cite{KurtFRFT}. In this case, the kurtosis is performed on the FRFT transformed signal representation. FRFT concentrates the energy of chirp signal while kurtosis can identify the peak energy concentration to detect the chirp signal. The authors propose to use FRFT filtering to filter out the detected component and repeat the detection process to iteratively detect multiple chirp components in the signal. The authors demonstrate the effectiveness of proposed approach by simulations. They adopt a threshold approach to detect the signal from white Gaussian noise (WGN) such that any signal with a kurtosis greater than zero will be considered a positive detection. However, in practical scenarios, the Gaussian distribution do not adequately model the underlying noise contributed by the receiver, surroundings etc \cite{practicalNWGN2,practicalNWGN}. Consequently, the kurtosis of signal deficient captures (noise) are non-zero values. Thus, an empirical approach is necessary to model the threshold to prevent false detections. Additionally, authors vaguely mention the matching order of FRFT is found by a one dimensional search procedure and do not reveal further information on the search. In a realistic unknown hostile jamming environment, FRFT matched orders of the jammers are unknown and the search procedure chosen to acquire the matching order is crucial to a successful detection and computational load. Furthermore, in this work, we also explore how the output from FRFT can be further exploited to perform localization of the RFI. 

Deploying algorithms on hardware platforms for commercial use is significantly challenging and requires careful consideration of several factors viz. robust performance, computational complexity, ease of use, relevant features of the prototype and meeting size-weight-and cost (SWaC) constraints.  There is a lack of comprehensive detection and analysis technique designed, and developed specifically for first responders to tackle intentional chirp signal that sweeps through a wide band and narrow band signal generated by unintentional RFI sources. Most approaches that solely rely on RFI signal energy deteriorate in low SNR scenarios. 
In this work, we propose a low cost, portable device, Jam-Guard, that performs robust RFI detection and localization designed specifically for Group-I, Group-II and Group-III jammers \cite{idad,jammergroups}. The utility of a RFI detector lies in its ability to detect very low power signals with exceptionally low false alarm rates. This requires adoption of a robust parallel detection module consisting of statistical techniques such as Kurtosis and time-frequency technique such as FRFT. To ensure lower complexity while ensuring high probability of detection, we adeptly employ Golden Section Search (GSS) \cite{gold} to efficiently converge at the FRFT matched order. Additionally, we adopt an FRFT based approach to localize using MUSIC \cite{music}. The matched order FRFT computed during detection is reused to prevent recomputation and save computational resources. Proposed detection outperforms the state-of-the-art  commercial RFI detectors. 
\begin{figure*}[h!]
\centering
\epsfig{file=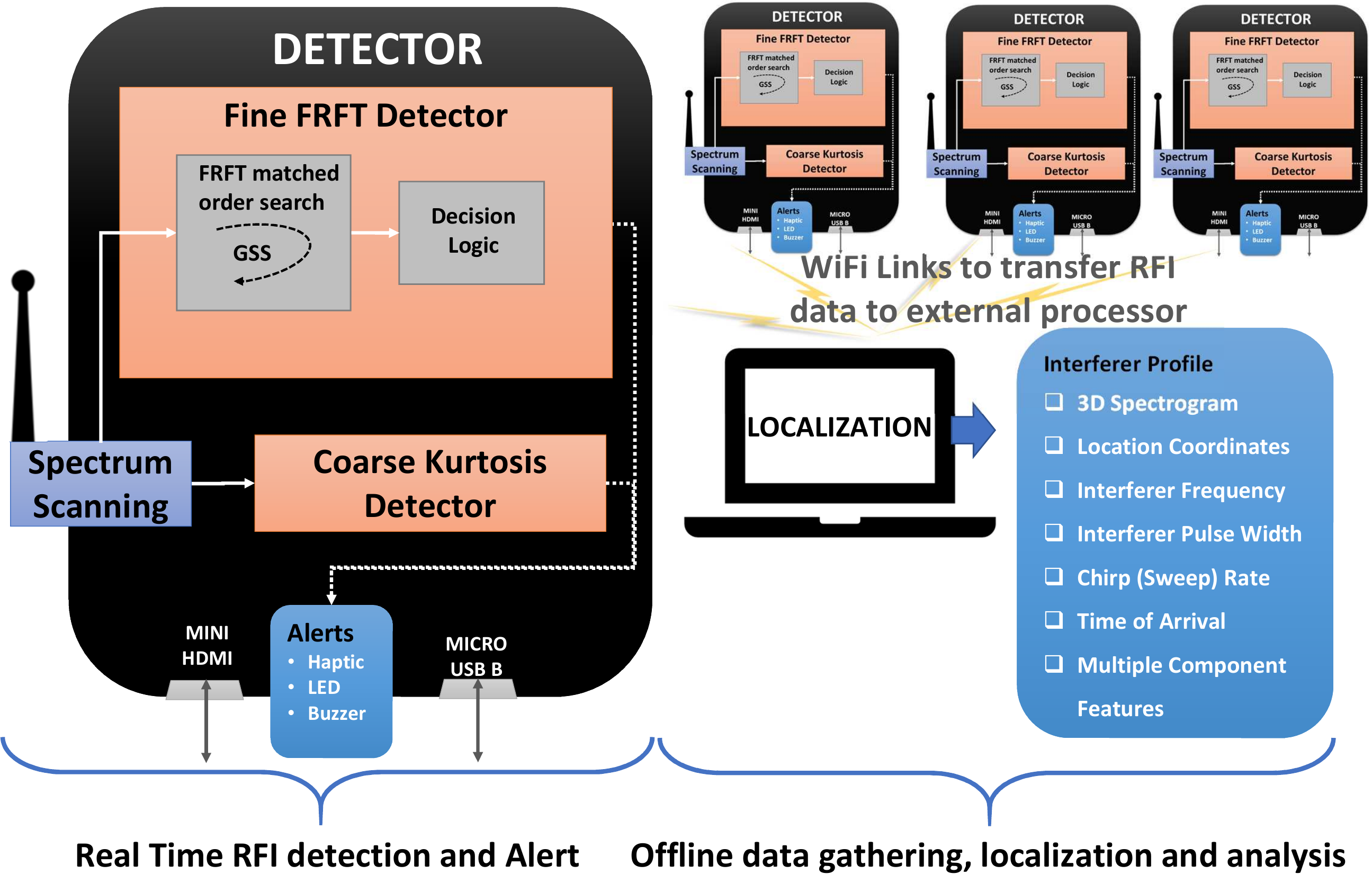, width=4.2 in,}
\caption{System Diagram of Proposed Device}\label{fig:IDAD}
\end{figure*}

Therefore, the contribution of this paper can be summarized as follows, 
\begin{itemize}
    \item To the best of our knowledge, this is the first work that proposes a comprehensive RFI detection and localization device while maintaining a low SWaC footprint.
    \item The FRFT-based parallel signal detection scheme is designed and preliminary feasibility is established using extensive experiments using SDRs.
    \item Output of FRFT-based detection scheme to perform localization has been established using simulations.
    \item Most importantly, Jam-Guard prototype is developed using Commercial Off-The-Shelf (COTS) components and the proposed algorithm is implemented and refined to establish a Pareto efficiency between computation load and accuracy of detection.  
\end{itemize}

\section{System Architecture}\label{sec:arch}

The overall system architecture of Jam-Guard is shown in Fig.\;\ref{fig:IDAD}. Each Jam-Guard consists of two parallel detection paths, (i) Coarse kurtosis based detector and (ii) Fine FRFT based detector. Each individual Jam-Guard also performs a preliminary analysis of the detected RFI and saves this information as a detection report locally. This information is reused during localization to prevent recomputing.

\subsection{Coarse Detector}
The coarse detection module performs kurtosis detection by considering the fourth and second moments of the frequency domain samples of the captured samples. Kurtosis is computed on the Fourier transform of the received Inphase-Quadrature (IQ) samples as follows,
\begin{align}
\mathfrak{K} = \frac{\frac{1}{N} \sum_{n=1}^{N}(\mathcal{R}(n) - \bar{\mathcal{R}})^4}{(\frac{1}{N} \sum_{n=1}^{N}(\mathcal{R}(n) - \bar{\mathcal{R}})^2)^2}
\end{align}
where $N$ is the number of IQ samples that are processed, $\mathcal{R}(n)$ denotes the absolute value of N-point FFT of the received samples and $\bar{\mathcal{R}}$ is the average FFT magnitude.
\subsection{Fine Detector}\label{sec:frft}
Fine detector houses an FRFT-order search procedure (GSS) to lock on to the matching FRFT order. RFI is detected if the matched order corresponding to a peak in the FRFT domain is obtained. To identify the peak efficiently, we compute kurtosis of the absolute value of FRFT. FRFT can give higher energy concentration of chirp signals in the FRFT domain once the transform order matches the chirp rate of the RFI. The GSS \cite{gold} in a closed interval searches for the order corresponding to the maximum FRFT value.  
The FRFT of the IQ samples are evaluated as per the digital computation proposed by Ozatkas et al. \cite{Ozatkas}.
\begin{align}
\mathfrak{F}^a[f(x)] = A_{\alpha}e^{j\pi\gamma x^2}\int_{-\infty}^{\infty}e^{j2\pi\beta xx'}[e^{j\pi\gamma x'^2}f(x')]dx',
\end{align}
where $\alpha=\frac{a\pi}{2}$ is the transform angle, $\gamma=\cot\alpha$, $\beta=\csc\alpha$ and $f(x)$ is the function that is being transformed. The discrete form of the transform requiring $O(N\log N)$ computations is expressed as 
\begin{align}
\mathfrak{F}^a[f(\frac{m}{2\Delta x})] = \frac{A_{\alpha}}{2\Delta x}e^{j\pi(\gamma-\beta) (m/2\Delta x)^2}\times \\ \notag 
\sum_{n=-N}^{N}e^{j\pi\beta ((m-n)/2\Delta x)^2}e^{j\pi(\gamma-\beta) (n/2\Delta x)^2}f(\frac{n}{2\Delta x}),
\end{align}
This expression can be realized with simpler operations such as chirp multiplication and convolution. The convolution can be performed in $O(N\log N)$ time by using FFT keeping the overall complexity at $O(N\log N)$. The computations are as per the detailed FRFT computation in \cite{frftcomp}. The time-frequency ($x-\nu$) plane of a signal can be viewed in the FRFT domain as the rotation of the $x-\nu$ axes around the origin making an angle $\alpha$ with the $x$-axis. Readers are encouraged to read the work by Ozatkas et al. \cite{Ozatkas} and A. Bultheel et al. \cite{frftcomp} to gain more insight into the FRFT computation.
The detection module logs the FRFT at the matched order, peak FRFT value, matched FRFT order, frequency bin corresponding to the peak FRFT value in the detection report.

\subsection{Localization-MUSIC}
In this section, we discuss how we leverage the matched order FRFT derived at detection phase to perform localization. We localize the wideband RFI source by modifying the Multiple Signal Classification (MUSIC) algorithm to obtain the direction of arrival (DoA). We consider a far field wideband Linear Frequency Modulated (LFM) chirp signal impinging a linear array of $M$ Jam-Guards on the $(x,y)$ plane with a direction of arrival ($\theta$).  The discrete $N-$point chirp signal as received by the $i^{th}$ Jam-Guard is expressed as 
\begin{equation}
r_i(t) = s(t)\nu_i(\theta) + n_i(t) \label{eq:rx}
\end{equation}

where $n_i(t)$ is the additive white Gaussian noise, $s(t) = e^{[j2\pi(ft + \frac{\mu}{2}t^2)]}$ is the chirp signal with initial frequency $f$ and chirp rate $\mu$, $\nu_i(\theta) = e^{\frac{-j(i-1)2\pi d\sin{\theta}}{\lambda}}$ is the  steering function of the $i-th$ Jam-Guard. The wideband signal is focused into a single frequency component by means of matched FRFT order ($a_{opt}$). This is attributed to the de-chirping effect of FRFT \cite{dechirp}. The matched-order FRFT can be expressed as, 
\begin{equation}
    \hat{r}_i(t) = \mathfrak{F}^{a_{opt}}[r_i(t)]
\end{equation}
Rewriting the equation (\ref{eq:rx}) in matrix form, we have
\begin{equation}
    \bold{R} = \bold{\Theta}s(t) + \bold{N}
\end{equation}
where $\bold{R}=[\hat{r}_1(t),\hat{r}_2(t), ..., \hat{r}_M(t)]^T$ is the array output data, $\bold{\Theta} = [1,  e^{\frac{-j2\pi d\sin{\theta}}{\lambda}}, ..., e^{\frac{-j(M-1)2\pi d\sin{\theta}}{\lambda}}]$, and $\bold{N}$ is the additive white Gaussian noise. Since saving computational resources is prominent for any real-time system performance, we reuse the matched order FRFT ($\mathfrak{F}^{a_{opt}}[r_i(t)]$) computed during the detection phase. The gist of MUSIC lies in decomposing the array output covariance matrix into signal and noise subspaces to form a spatial spectrum function. The array output covariance matrix can be obtained by sample-averaging the array output data as $\bold{R}_{cov} = (1/N)\bold{R}\bold{R}^H$. Eigen decomposing the array output covariance matrix separates it into signal and noise subspaces. After arranging the eigenvalue-eigen vectors in the descending order, the first column corresponds to the signal subspace while the remaining $(M-1)$ corresponds to noise subspace ($\bold{R}_n$). To account for low SNR scenarios, an improved-MUSIC \cite{imusic} is considered which performs spatial smoothing of $\bold{R}_{cov}$ such that $\bold{R}_{c} = \bold{R}_{cov} + \bold{T}\bold{R}_{cov}^*\bold{T}$, where $\bold{T}=\mathfrak{f}[\mathfrak{I}(M)]$ is the transition matrix and $*$ is the complex conjugate. Here, the function $\mathfrak{f}[\bold{X}]$ flips the columns of matrix $\bold{X}$ from left to right and $\mathfrak{I}(M)$ creates an identity matrix of size $M$ with ones on its main diagonal and zeros elsewhere. Thus, new noise subspace $\bold{R}_{cn}$ is computed for the transitioned covariance matrix $\bold{R}_{c}$. The spatial spectrum function takes the form,
\begin{equation}
    \mathcal{M}(\theta) = \frac{1}{\Theta^H\bold{R}_{cn}\bold{R}_{cn}^H \Theta}
\end{equation}

A one-dimensional search for the direction of arrivals (azimuths) along the direction of maximum spatial spectrum will return the true azimuth. 

\section{Preliminary Feasibility Analysis}\label{sec:simresults}
The preliminary goal of this work was to conduct a rapid feasibility analysis of the proposed detector framework in terms probability of detection ($P_d$) under various operating scenarios. In this regard, we have used Universal Software Radio Peripheral (USRP), a software defined radio, to generate few generic RFI waveforms to be transmitted in the  industrial, scientific and medical (ISM) radio bands  to emulate the behaviors of jammers for the purpose of OTA evaluation. The waveforms were developed on GNU Radio platform (open-source signal processing software). The signal processing for generating the waveforms was written in Python and imported into GNURadio. An AirSpy mini SDR \cite{airspy} connected to the host laptop was acting as the receiver. AirSpy mini can be programmed with the \emph{libairspy} application programming interface. The detection algorithms were run on a Linux laptop with Intel i5-4670 processor.

\begin{figure}[h!]
\centering
\epsfig{file=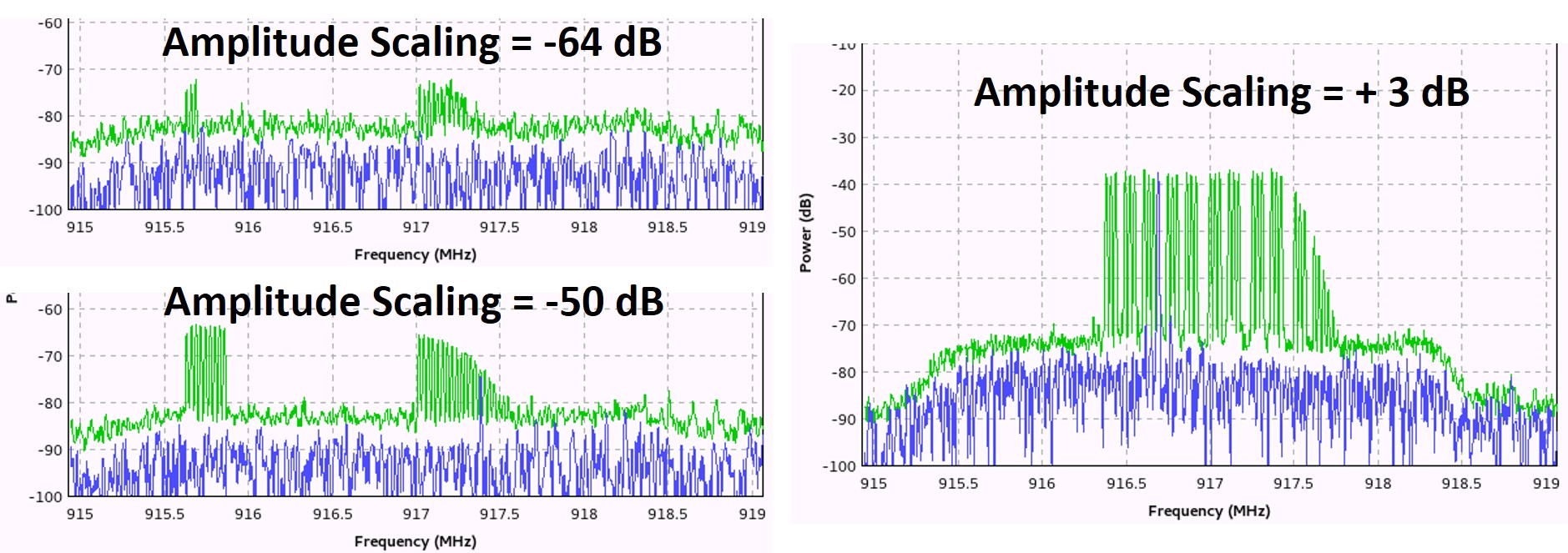, width=3.4 in,}
\caption{SNR at different Jammer Scaling level}\label{fig:RFI}
\end{figure}

To evaluate the performance of the detector (implemented in C++) under varying SNR, we applied digital scaling factor to the generated RFI. The varying RFI amplitudes are represented as varying scaling factor using a dB scale. In order to visualize the operating conditions, three power spectrum plots of the RFI received at different scaling factors are shown in Fig.\;\ref{fig:RFI}. The green color depicts the peak hold function at each frequency, whereas the blue color shows the instantaneous signal energy. It is important to realize that the instantaneous SNR is lower than the impression that the peak hold plot provides. 

\begin{figure}[h!]
\centering
\epsfig{file=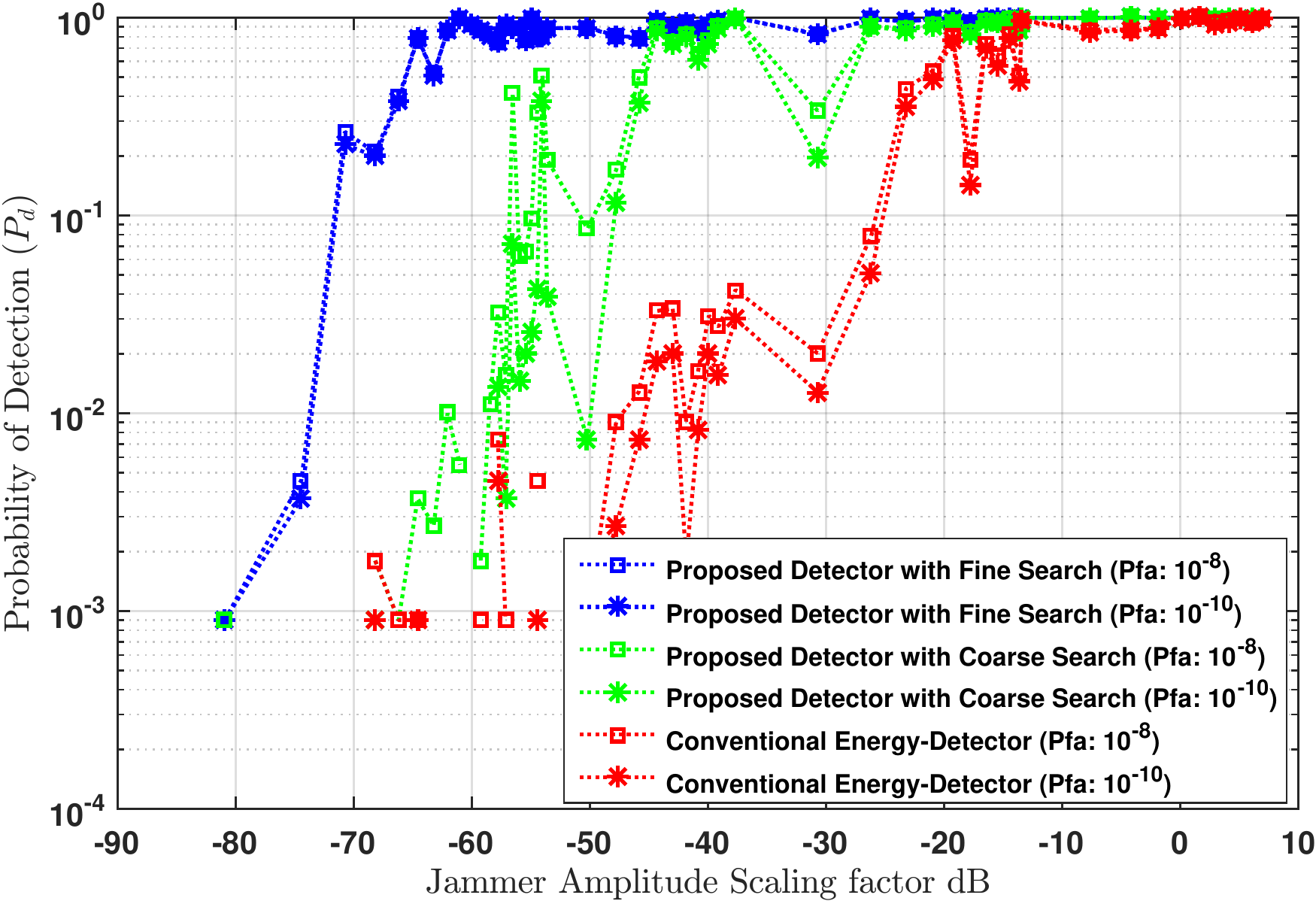, width=3.4 in,}
\caption{Probability of Detection vs Signal Strength}\label{fig:Result_sim}
\end{figure}

The RFI generated is a Group-II LFM chirp waveform sweeping at a rate of $2\;\mathrm{MHz/s}$. The AirSpy mini (SDR) is tuned to listen to $6\;\mathrm{MHz}$ bandwidth with center frequency at $917\;\mathrm{MHz}$. The performance is evaluated for three detectors, (i) conventional energy detector, (ii) proposed detector with coarse search and (iii) proposed detector with fine search. The coarse search performs a simple three order search at predetermined FRFT orders $[ 0.2,0.3,0.4]$ and used mean FRFT magnitude as the test statistic. The fine search employed the GSS algorithm with convergence rate $10^{-8}$ to determine the matched order and uses kurtosis of FRFT as mentioned in subsection \ref{sec:frft}. For each of these detectors, we evaluate threshold set accordingly for two false alarm rates $P_{fa}=[10^{-8},10^{-10}]$. Each data point in Fig. \ref{fig:Result_sim} corresponds to an average of $1095$ spectral snapshots with receiver sampling at a rate of $3\;\mathrm{MS/s}$. At each snapshot, the detector processes 4096 IQ samples.
 At very low SNR scenario (i.e. when Jammer scaling factor was set to $-60\;\mathrm{dB}$ ), the proposed solution with GSS renders a significant improvement as opposed to coarse search. The proposed detector with fine search starts to provide near absolute detection even at low SNR conditions which further saturates with increasing RFI power. Both versions of the proposed detector outperforms conventional energy detector which delivers acceptable detection rates only when the RFI power is high enough (i.e. when Jammer scaling factor was set to ~$-20\;\mathrm{dB}$ to ~$-10\;\mathrm{dB}$).  Therefore, our preliminary analysis has shown that the proposed device will significantly outperform (by up to ~$40\;\mathrm{dB}$) devices that depend on conventional energy detector at a given $P_{fa}$.
 
\begin{figure}[h!]
\centering
\epsfig{file=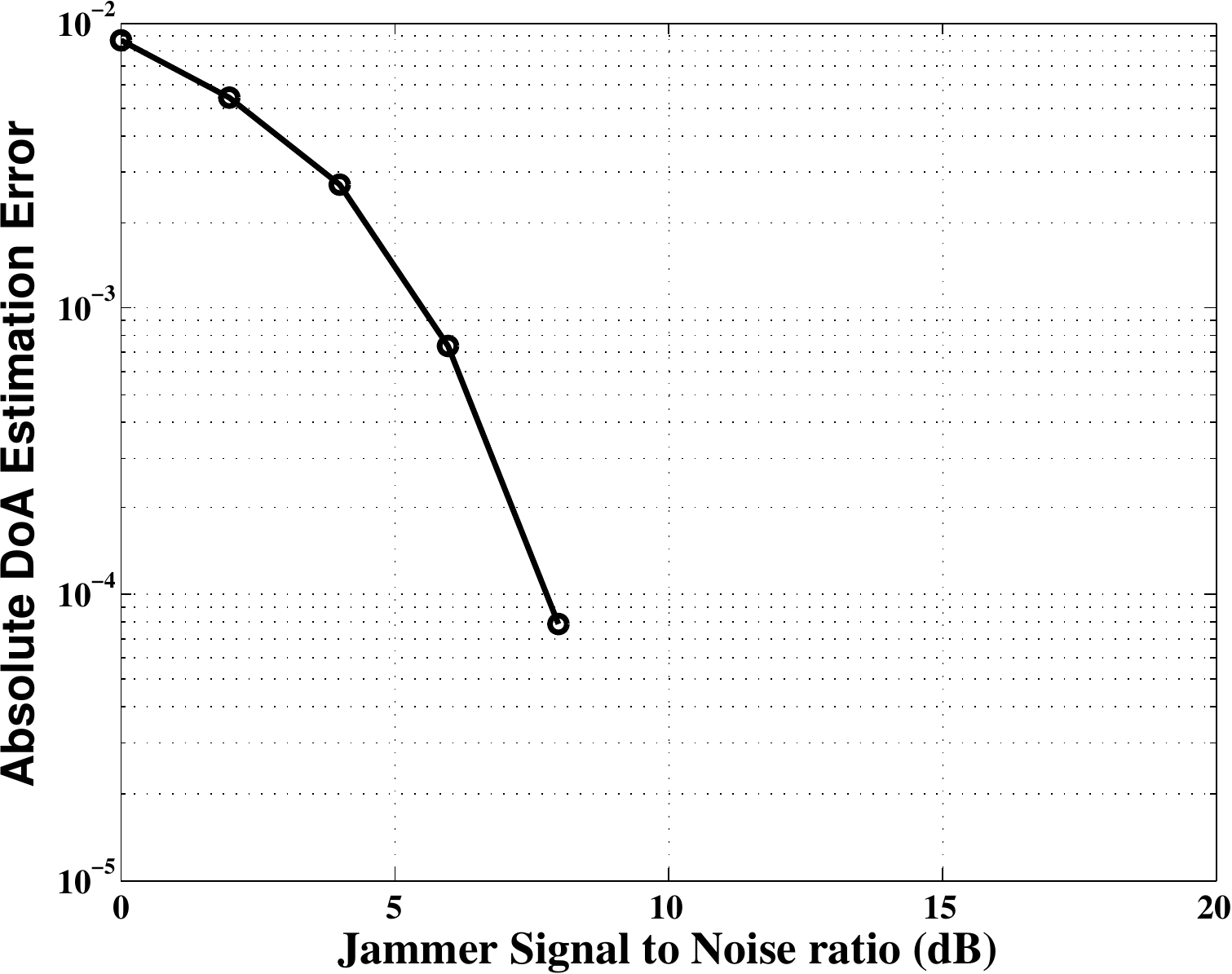, width=3 in,}
\caption{Absolute Azimuth estimation error vs Signal Strength}\label{fig:doa}

\end{figure}

Further, to test the performance of the improved-MUSIC approach, we perform computer simulations in MATLAB. For the test scenario, we consider a LFM source incident at an azimuth of $20^{\text{o}}$ upon two Jam-Guards located on a first responder vehicle at a distance $d=\lambda/2$ apart. To study the effect of varying RFI power, we vary the SNR of incident LFM from $0$ to $20$ dB in steps of $2$ dB. Figure~\ref{fig:doa} plots the absolute azimuth estimation error for varying SNR. The azimuth estimation error degrades with increasing SNR.

The algorithmic effectiveness of the proposed detection scheme was successfully established from these preliminary tests. Bear in mind, these were run on an Intel i5 processor. The same tests would be slower on any low cost embedded platform. Once of the key challenges of this work is to develop superior RFI detection framework on the embedded platform owing to the SWaC constraints. In the next section, we will conduct an extensive complexity analysis which forms the backbone of the Jam-Guard prototype development.

\section{Prototype Development and Evaluation}\label{sec:protoeval}

In our design, the main development board is a Raspberry Pi (RPI) 3 Model B with Quad Core Broadcom BCM2837 64-bit ARMv8 processor and weighing $41.2\;\mathrm{g}$. The choice was motivated from the low cost, size and large open community support for RPI development. Additionally, it is enabled with WiFi (802.11 b/g/n). To reside within the SWaC constraints of the device, Jam-Guard uses AirSpy mini which is compatible with RPI and weighs $50\;\mathrm{g}$. The technical specifications include $3.5$ dB NF between $42$ and $1002\;\mathrm{MHz}$, $12$ bit ADC at $20$ MSPS and $10$, $6$ and $3$ MSPS IQ output. 

\begin{figure}[h!]
\centering
\epsfig{file=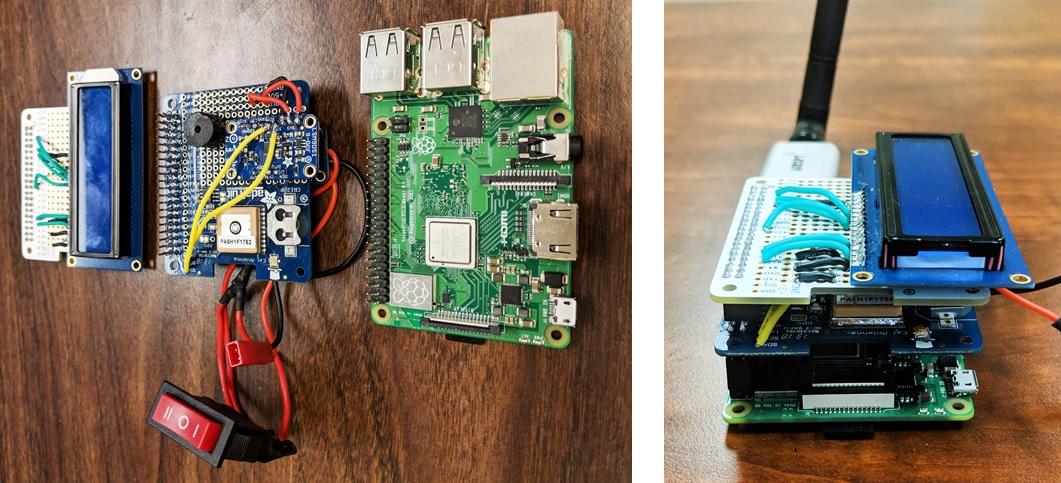, width=3.4 in,}
\caption{Component layout of Jam-Guard}\label{fig:Component}
\end{figure}

\begin{figure}[h!]
\centering
\epsfig{file=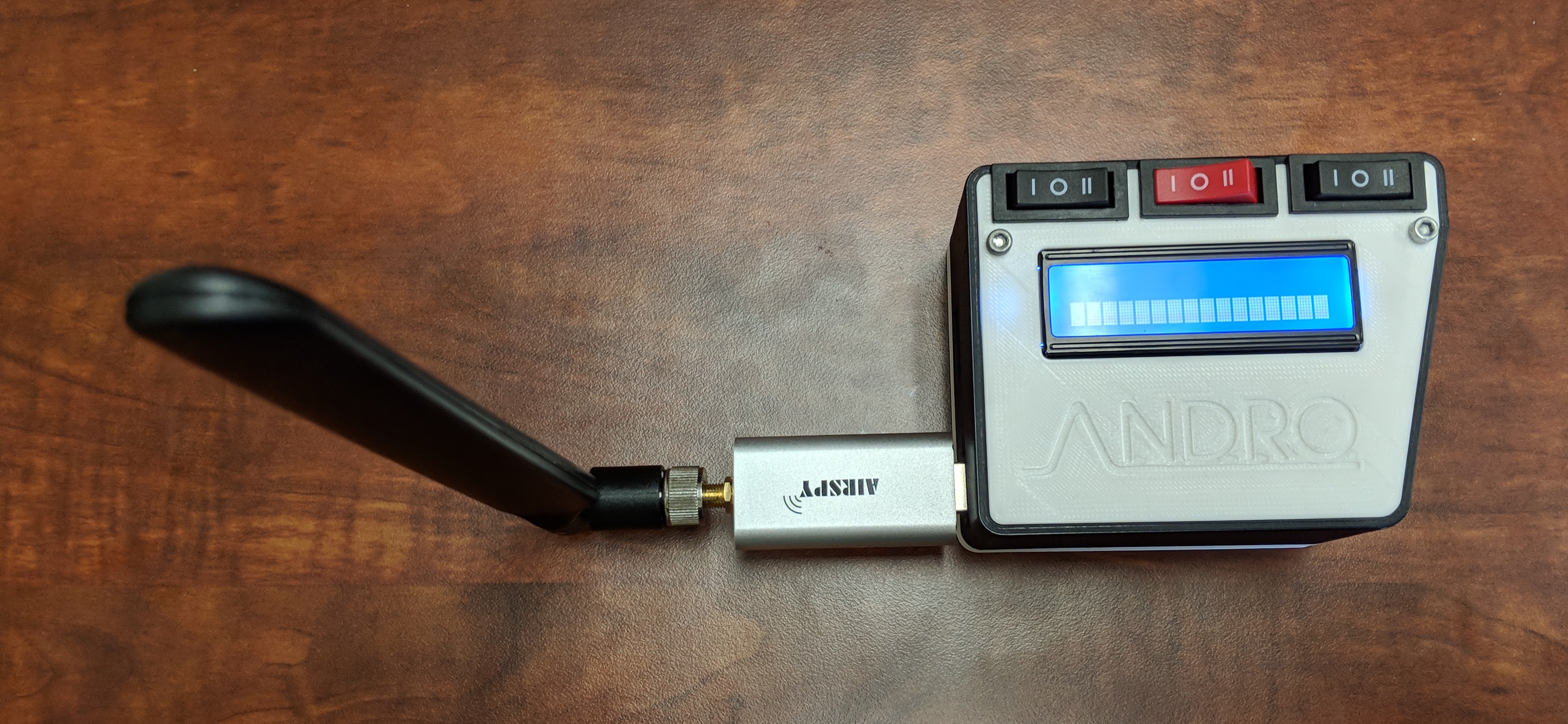, width=3.2 in,}
\caption{Jam-Guard}\label{fig:Prototype}
\end{figure}

To generate three types of feedback, Jam-Guard possess; (i) Mini LCD screen: provide visual feedback by displaying basic information of the detected RFI, (ii) Audio buzzer: beeps to indicate the detection of RFI in the given channel and (iii) Haptic motor controller: provide haptic feedback to the end users. Jam-Guard further includes Perma-Proto HAT, Ultimate GPS HAT, Stacking Header, $3.7V$ Lithium Ion battery pack, a nine degree of freedom inertial measurement unit (IMU) sensor (3-axis accelerometer, 3-axis magnetometer and a 3-axis gyroscope) to estimate location when GPS is temporarily unavailable, and three 3-pin switches. The demystified layout and the complete prototype of Jam-Guard are shown in Fig.\;\ref{fig:Component} and Fig.~\ref{fig:Prototype} respectively. 

\subsubsection{Complexity Analysis}
The algorithmic feasibility was established in the previous section. To achieve real-time performance from Jam-Guard, it is imperative to establish computational feasibility on the target embedded platform. In order to emphasize the significance of minimizing the CPU load or in other words to hasten the detection process, in this section, we will discuss the computational complexity of adopting proposed solution and how the current prototype of Jam-Guard ensures rapid detection.

In its early phases of development, Jam-Guard utilized an iterative search procedure which would search for the matching FRFT order in a closed interval $[0.001,1.51]$ that yielded a detection probability of $0.947$. But this approach was impractical for a real-time detection for first responders as it was rather slower and required $120\times10^8$ CPU cycles to complete which would imply more than an hour to detect on an intel i5 processor. The need for a faster search procedure formed the motivation behind adopting GSS. The GSS approach with a convergence rate of $10^{-8}$ and $10^{-3}$ drastically reduced the CPU cycles to $30\times 10^8$ and $10\times10^8$ respectively while yielding a detection probability of $0.9973$. This implied an execution time in the order of several seconds on the RPI platform which where impractical and required to be further reduced to deliver real-time performance. To further speed up the detection, the FRFT computation had to be sped up such that all factors contributing to the detection time are covered. 

Subsequently, a faster implementation of FRFT was adopted. The faster approach followed a two-phase implementation as opposed to the single-phase approach by \cite{frftcomp}. The two-phase approach splits the samples into even and odd parts and performs all operations on each of them separately such that only the necessary samples are computed \cite{poly}. We followed this two-phase approach along with the GSS of convergence rate $10^{-3}$ to arrive at a reduced CPU cycles of $5\times10^8$ with a detection probability of $0.9973$. This implied a faster detection time of ~$\leq 1$ s on RPI platform. The complexity reduction is summarized in table \ref{tab:complexity}. The higher detection probability as opposed to iterative search can be attributed to the finer FRFT order resolution of the GSS realized by the Golden ratio. The resolution of iterative search is $0.01$ which is very coarse compared to GSS. Reducing the step size of iterative search will only increase the detection time and hence not ideal for the application at hand.

\begin{table}[h!]
    \centering
    \caption{Complexity Analysis of Jam-Guard Detection Schemes\label{tab:complexity}}{%
    \begin{tabular}{p{4.1 cm}p{1.4 cm}p{0.7 cm}} 
    \toprule
      \textbf{Jam-Guard Detection Schemes}   & \textbf{CPU Cycles}  & $\mathbf{P_d}$\\ 
      \midrule
      Iterative Search $[0.001,1.51]$.   & $120\times10^8$ &  0.947   \\ 
      GSS convergence rate $10^{-8}$   &   $30\times10^8$ &  0.9973   \\ 
      GSS convergence rate $10^{-3}$   &   $17\times10^8$ &  0.9973   \\ 
      GSS ($10^{-3}$) with Two-phase FRFT    & $  6\times10^8$ &  0.9973   \\ \bottomrule
    \end{tabular}}{}
\end{table}
\subsubsection{Prototype Performance Analysis}

The entire detection is implemented on the RPI platform in C++. The detection algorithm hosts two boost threads to acquire the samples from AirSpy mini receiver and process them for detection. The coarse and fine detection processes run in parallel and trigger the GPIO pins of the RPI to trigger the alerts upon RFI detection. A pipe is employed to perform sanity checks between the two parallel processes.

\begin{figure}[h!]
\centering
\epsfig{file=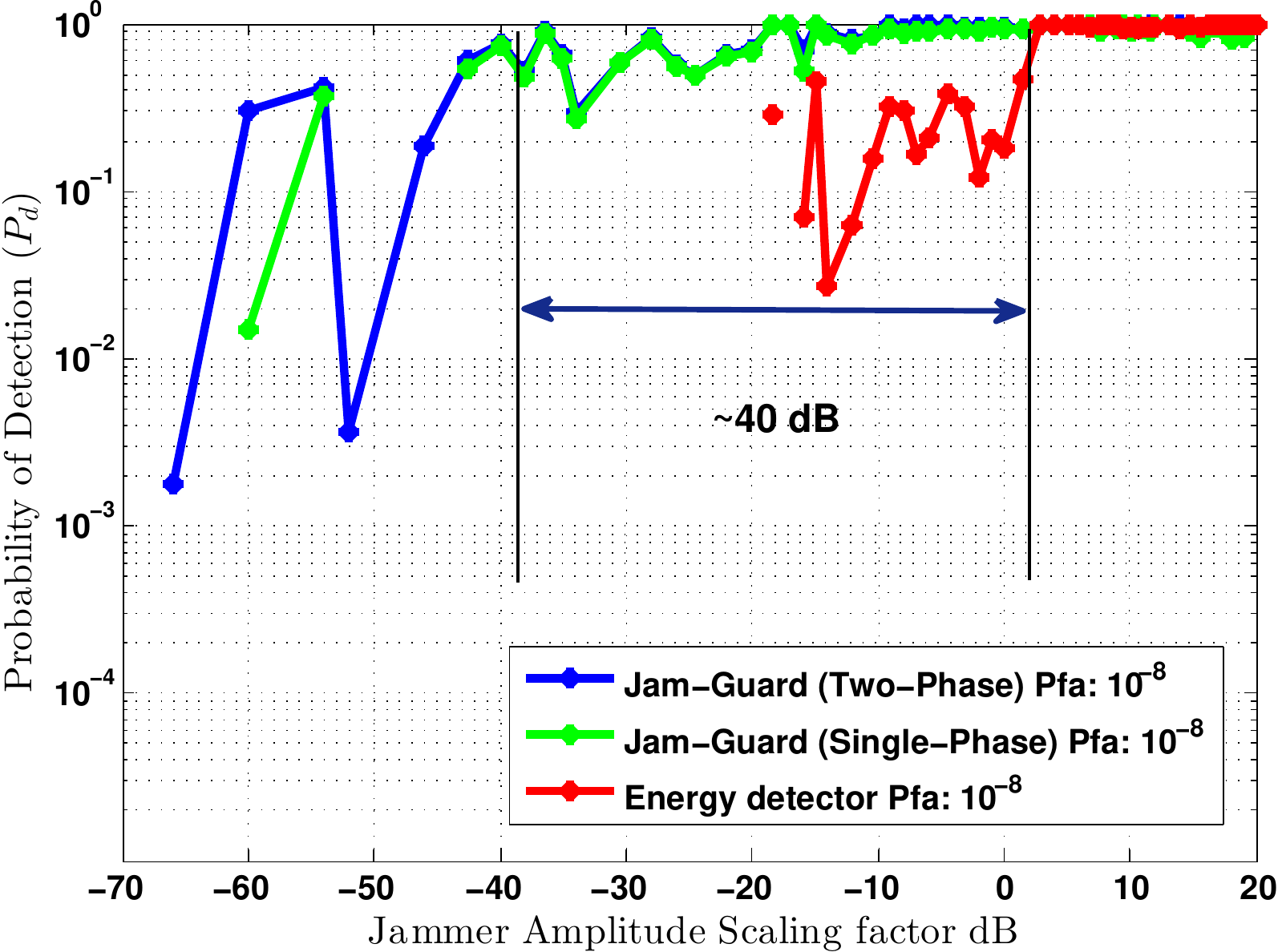, width=3.4 in,}
\caption{Probability of Detection vs Signal Strength}\label{fig:Result}
\end{figure}

In order to illustrate the effectiveness of Jam-Guard compared to conventional Energy detectors on the embedded platform, we perform real-time testing of the detection algorithm running on Jam-Guard under three settings; single-phase FRFT with GSS convergence $10^{-3}$ (green curve), two-phase FRFT with GSS convergence $10^{-3}$ (blue curve), and conventional energy detection (red curve). Jam-Guard receives the samples at a sampling rate of $3$ MS/s and captures a $0.34$ ms snapshot at a center frequency of $917$ MHz. The RFI settings are the same as in the preliminary testing. The receiver operating characteristics (ROC) curve of Jam-Guard once again showcases superior detection capability as opposed to conventional Energy detectors as shown in Fig.~\ref{fig:Result}. The two-phase FRFT curve performs as good as the single-phase FRFT while achieving CPU performance gain of $\sim 66\%$. A notable feature of using FRFT as opposed to other cyclostationary approaches is the ability to detect with very few samples ($1024$ samples) which significantly saves the CPU load.

\section{Conclusion and Future Work}\label{sec:con}

This work showcased a detailed account of the development cycle and evaluations of Jam-Guard that significantly outperforms the state-of-the-art devices in market. We have presented a detailed implementation analysis of Jam-Guard, hardware details, and presented the in-depth preliminary and embedded platform experimental results. The low SWaC of the proposed device will enable first responders and other users to carry it with ease during their mission. Jam-Guard is equipped with three types of alerts; audio, visual and haptic to alert the users of the detected RFI thereby improving efficiency and safety. The future work involves developing a dual antenna version of Jam-Guard, virtual array techniques to localize more than one RFI source, and two-dimensional DoA estimation such that the azimuth and elevation of the RFI source can be deduced by a single Jam-Guard. 

\bibliographystyle{ieeetr}
\bibliography{IDAD}
\end{document}